\pdfoutput=1

\documentclass[11pt]{article}

\usepackage[]{acl}

\usepackage{times}
\usepackage{latexsym}
\usepackage{multirow}
\usepackage{booktabs} 
\usepackage[T1]{fontenc}

\usepackage[utf8]{inputenc}
\usepackage{subcaption}
\usepackage{arydshln}
\usepackage{microtype}

\usepackage{inconsolata}

\usepackage{graphicx}
\usepackage{xcolor}
\usepackage{url}
\definecolor{red}{rgb}{0.99, 0.02, 0.02}
\NewDocumentCommand{\heng}
{ mO{} }{\textcolor{red}{\textsuperscript{\textit{Heng}}\textsf{\textbf{\small[#1]}}}}
\NewDocumentCommand{\revanth}
{ mO{} }{\textcolor{blue}{\textsuperscript{\textit{Revanth}}\textsf{\textbf{\small[#1]}}}}
\NewDocumentCommand{\ray}
{ mO{} }{\textcolor{orange}{\textsuperscript{\textit{ray}}\textsf{\textbf{\small[#1]}}}}
\newcommand{\name}{\textsc{FIRST}}

\title{\name{}: Faster Improved Listwise Reranking with Single Token Decoding}

\author{Revanth Gangi Reddy$^1$\thanks{Equal Contribution.}\hspace{0.15em}JaeHyeok Doo$^{1,2}$\footnotemark[1]\hspace{0.15em}Yifei Xu$^{1,2}$\footnotemark[1]\hspace{0.15em}Md Arafat Sultan$^3$\hspace{0.3em}Deevya Swain$^{1,2}$\\\textbf{Avirup Sil}$^3$\hspace{1em}\textbf{Heng Ji}$^1$\hspace{1em}\\
$^1$University of Illinois Urbana-Champaign \hspace{1em} $^2$Lapis Labs \hspace{1em} $^3$IBM Research AI \\
  \texttt{\{revanth3,jdoo2,yifeix5,deevyas2,hengji\}@illinois.edu}\\ \texttt{arafat.sultan@ibm.com} \hspace{1em} \texttt{avi@us.ibm.com}\\
  }

\begin{document}
\maketitle
\begin{abstract}

Large Language Models (LLMs) have significantly advanced the field of information retrieval, particularly for reranking. Listwise LLM rerankers have showcased superior performance and generalizability compared to existing supervised approaches. However, conventional listwise LLM reranking methods lack efficiency as they provide ranking output in the form of a generated ordered sequence of candidate passage identifiers. Further, they are trained with the typical language modeling objective, which treats all ranking errors uniformly--potentially at the cost of misranking highly relevant passages. Addressing these limitations, we introduce \name{}\footnote{\url{https://github.com/gangiswag/llm-reranker}}, a novel listwise LLM reranking approach leveraging the output logits of the first generated 
identifier to directly obtain a ranked ordering of the candidates. Further, we incorporate a learning-to-rank loss during training, prioritizing ranking accuracy for the more relevant passages. Empirical results demonstrate that \name{} accelerates inference by 50\% while maintaining a robust ranking performance with gains across the BEIR benchmark. Finally, to illustrate the practical effectiveness of listwise LLM rerankers, we investigate their application in providing relevance feedback for retrievers during inference. Our results show that LLM rerankers can provide a stronger distillation signal compared to cross-encoders, yielding substantial improvements in retriever recall after relevance feedback. 

\end{abstract}

\section{Introduction}
\label{sec:introduction}

\begin{figure}[t]
    \centering
    \includegraphics[width=0.925\linewidth]{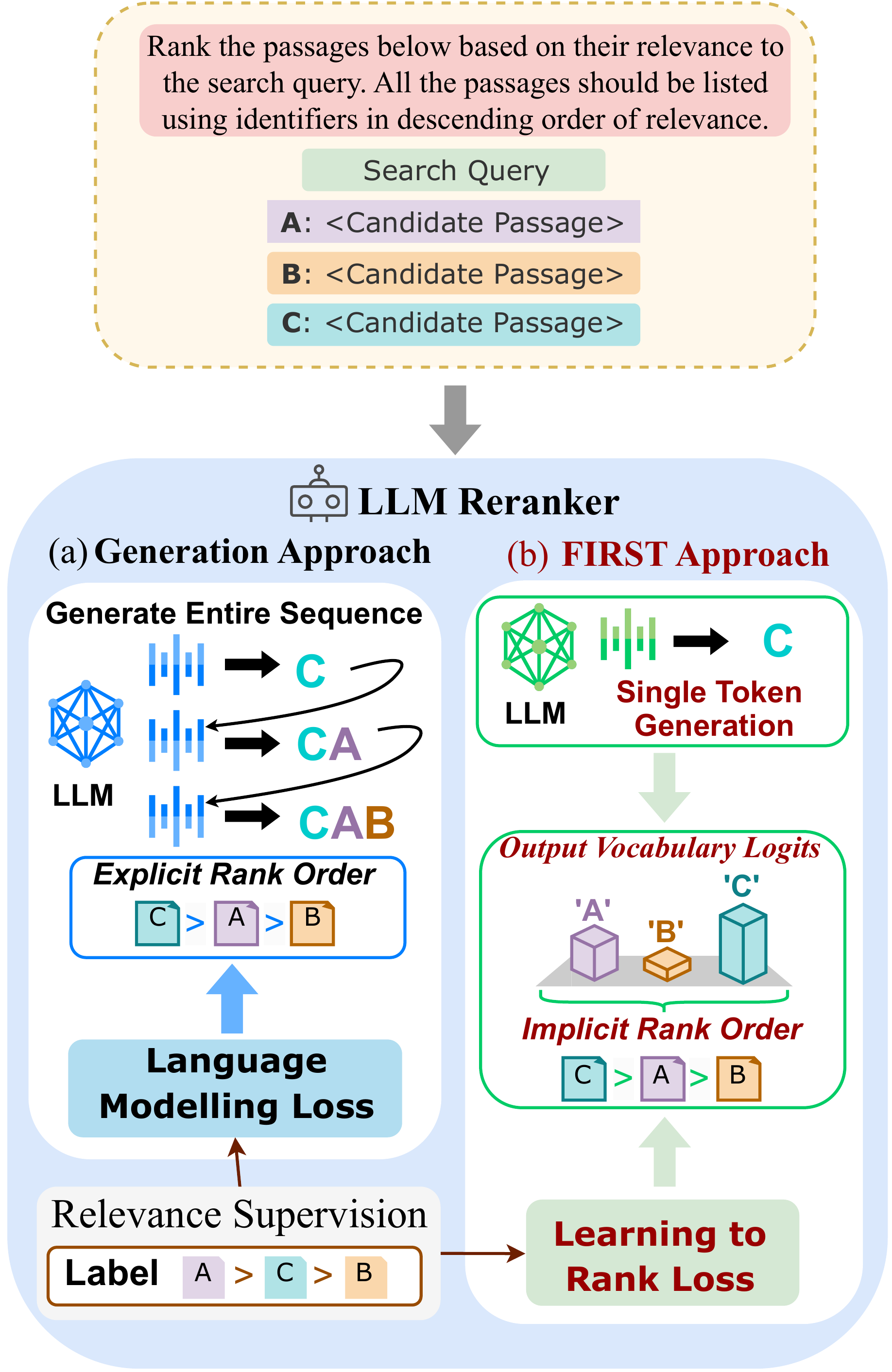}
    \vspace{-0.5em}
    \caption{\name{} (b) directly ranks candidates using the output vocabulary logits for the first generated identifier, as opposed to the generation approach (a) of generating the entire ordered sequence. A learning-to-rank loss is incorporated during training to provide supervision to the model for ranking using single-token decoding.}
    \label{fig:method}
\end{figure}

Given their vast linguistic knowledge and strong zero-shot capabilities~\cite{wei2022emergent}, there has been a natural push to incorporate large language models (LLMs) into the search stack~\cite{zhu2023large, wang2024large}.
One of the core applications of LLMs in search involves ranking candidate passages for their relevance to a given query.
Recent studies~\cite{sun2023chatgpt} have shown that instruction-tuned LLMs can outperform traditional supervised cross-encoders in zero-shot passage reranking~\cite{nogueira2020document, zhuang2023rankt5}.
In particular, listwise reranking approaches~\cite{tang2023found, pradeep2023rankzephyr} have received increased attention for their ability to score multiple passages simultaneously, as opposed to pointwise~\cite{zhuang2023beyond, zhuang2023open} or pairwise~\cite{qin2023large} reranking, where scoring is performed in isolation.
As~\citet{xian2023learning} have demonstrated, listwise reranking benefits from contextually comparing multiple passages at once, which helps calibrate relevance scoring better.

Listwise reranking with LLMs is typically framed as a generation task, where given a query and multiple candidate passages as input, the model outputs a ranked sequence of passage IDs. 
While~\citet{sun2023chatgpt, ma2023zero} use proprietary models,~\citet{pradeep2023rankvicuna, pradeep2023rankzephyr} demonstrate that open-source LLMs finetuned with GPT-3.5/GPT-4~\cite{achiam2023gpt} annotated data can also achieve competitive performance. 
~\citet{pradeep2023rankzephyr} introduce RankZephyr, which is trained using a standard language modeling objective, with the ranking sequence generated by GPT-4 as the target.
While this approach has shown promise, it has a number of key drawbacks.
First, it involves generating entire sequences of passage IDs, which is arguably inefficient, and as we demonstrate through our study, is also unnecessary.
Second, it penalizes errors uniformly across the ranking sequence; misjudging the rank of the most (and potentially only) relevant passage, for example, receives the same penalty as incorrectly swapping the ranks of two non-relevant passages.
Intuitively, reranker training should prioritize accurately ranking top candidates over those that bear low relevance to the query.

\begin{figure}[t]
    \centering
    \includegraphics[scale=0.7]{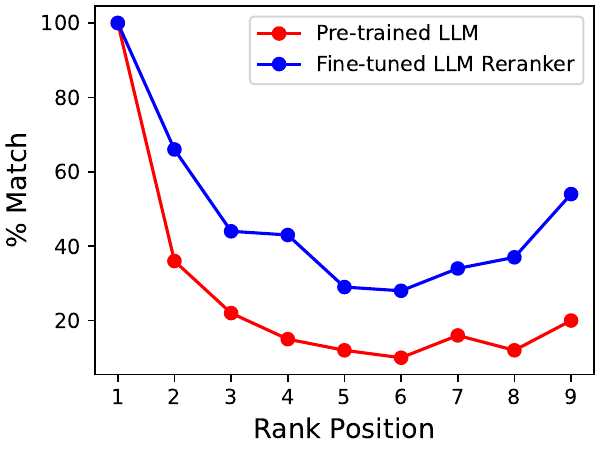}
    \caption{The $\%$ of times the rank \textit{generated} by an LLM reranker (RankZephyr~\cite{pradeep2023rankzephyr}) for a candidate agrees with the rank \textit{implied by its computed logit} for the same candidate in the first (top-rank) token position, at different ranks. 
    RankZephyr, originally fine-tuned with a sequence generation objective (in blue), shows a considerably higher similarity between the two above rankings than a pretrained LLM (in red).}
    \label{fig:ranking_matching}
\end{figure}

The goal of this work is to enable LLM rerankers to overcome these limitations.
Our investigation starts with the following question: Do the logits computed by existing LLM rerankers for their \textit{first} generated identifier, which are meant to only predict the top-ranked candidate, also provide a calibrated estimate of the relative importance of \textit{all} the input candidates?
In Figure~\ref{fig:ranking_matching}, we show how the ranking indicated by the logits produced by RankZephyr~\cite{pradeep2023rankzephyr} in its first token position matches that of its fully generated ranking sequence. 
We observe that RankZephyr's sequence-generation training objective also improves the quality of its logit-induced ranking by bringing it closer to the sequence-based ranking.
Crucially, this suggests that LLM rerankers can implicitly judge the relevance of candidate passages without needing to explicitly generate a ranking sequence.
We seek to capitalize on this property to significantly accelerate their inference process for listwise ranking, eliminating the need to generate a full sequence of IDs.

To that end, we present \name{}\footnote{\textbf{F}aster \textbf{I}mproved \textbf{R}e-ranking with a \textbf{S}ingle \textbf{T}oken}, a novel approach that relies solely on the output logits of the first generated identifier to produce a listwise ranking of input candidates.
\name{} employs a novel training strategy that directly incorporates a ranking loss into the supervision of LLM rerankers.
The use of a learning-to-rank loss~\cite{liu2009learning} also enables us to assign greater weights to important ranks, unlike generation-based losses that treat all ranks in the output sequence uniformly.
Figure \ref{fig:method} illustrates \name{}, our proposed approach for listwise LLM reranking. 
Single-token decoding not only improves the efficiency of inference but also maintains high performance by leveraging the more effective learning-to-rank supervision during training. 
Experiments in \S{\ref{sec:latency}} demonstrate that \name{} lowers the latency of LLM rerankers by 50\%.

We further demonstrate the benefits of \name{} in downstream applications.
Specifically, we study the impact of using LLM rerankers for pseudo-relevance feedback~\cite{rocchio1971relevance}, wherein the output of a reranker is used to improve the retriever recall at inference. Prior work~\cite{reddy2023inference, sung2023optimizing} typically uses numeric point-wise scoring output from cross-encoders~\cite{thakur-2020-AugSBERT} as the distillation supervision for relevance feedback. Here, we demonstrate (in \S{\ref{sec:relevance_feedback}}) that a superior output from an LLM reranker, although in the form of an ordering sequence, can provide better relevance feedback that leads to greater improvement in retriever recall when distilled with ranking losses.




The main contributions of this work are:
\begin{itemize}
    \item We introduce \name{}, a novel strategy for reranking with LLMs that obtains the ranking from only the output logits of the first generated identifier.
    \item By incorporating a learning-to-rank loss for supervision, \name{} improves ranking performance while lowering latency of inference by 50\%. 
    \item Finally, we demonstrate the potential of LLM rerankers for relevance feedback, with improved retriever recall compared to using cross-encoders for inference-time distillation.
\end{itemize}

\section{Related Work }
\label{sec:related_work}
\subsection{Reranking with LLMs}
Modern IR systems commonly employ a multi-stage pipeline, wherein an efficient initial retriever~\cite{robertson2009probabilistic, karpukhin2020dense} selects a set of candidates from a vast corpus, which is then reranked by a more sophisticated reranker~\cite{nogueira2019passage, nogueira2020document} to enhance precision. Methods leveraging cross-encoder models~\cite{nogueira2020document, zhuang2023rankt5} for rerankers have achieved notable success in improving ranking performance. Nonetheless, a principal limitation of such methodologies is their reliance on extensive in-domain human supervision, which leads to poor generalizability across different domains~\cite{zhu2023large}. Recent efforts have explored mitigating this limitation by utilizing the zero-shot capabilities of LLMs for passage reranking \cite{ma2023zero, sun2023chatgpt}. Building on this,~\citet{pradeep2023rankvicuna, pradeep2023rankzephyr} finetuned open-source LLMs to be capable of performing high-quality listwise reranking on par with proprietary models, such as GPT-4~\cite{achiam2023gpt}. 
However, existing works do not incorporate any traditional learning-to-rank strategies~\cite{liu2009learning} when finetuning LLMs for listwise reranking. Further, they often overlook the considerable latency of reranking with LLMs.
Our approach, \name{}, addresses both limitations by leveraging the output logits of the first generated identifier to directly obtain the rank order. \name{} successfully demonstrates that substantial efficiency gains are achievable without compromising accuracy in reranking with LLMs.

\subsection{Learning to Rank}
In IR literature, Learning to Rank (LTR)~\cite{liu2009learning} aims to order items by their relevance to a particular query. LTR is an extensively explored research field, and multiple optimization techniques have been proposed that can be broadly categorized into three main approaches: pointwise, pairwise, and listwise. Given the item and query pair, pointwise approaches~\cite{crammer2001pranking, li2007mcrank} determine relevance by a numerical score or binary judgment, which is later used for ranking. The pairwise approaches~\cite{burges2005ranknet, burges2006lambdarank} measure the pairwise preferences between item pairs, being reportedly more effective than the pointwise method by capturing the relative importance of the items. Later, the training subjects were extended to a list of items, and the loss was defined over the entire item list~\cite{cao2007listnet, xia2008listmle, taylor2008softrank}, allowing to obtain more fine-grained relative importance among the items. Recent studies~\cite{nogueira2020document, zhuang2023rankt5, sun2023chatgpt, pradeep2023rankvicuna, pradeep2023rankzephyr} have applied pre-trained language models for passage reranking and observed significant performance gains. 
While ~\citet{zhuang2023rankt5} and ~\citet{sun2023chatgpt} employ LTR algorithms for finetuning, they only consider it for pointwise ranking. On the other hand, our approach adopts LTR algorithms for finetuning listwise LLM rerankers.

\subsection{Listwise Reranking}
Early exploration of leveraging pre-trained language models for document reranking relied on pointwise ranking~\cite{sachan2022improving, cho2023discrete, zhuang2023rankt5}. This involves extracting the generation probability of a relevance token, such as `true' or `yes', from the model when asked to determine the document's relevance to a query. Despite their supremacy over supervised ranking methods based on cross-encoders~\cite{nogueira2020document, zhuang2023rankt5}, the isolated scoring mechanism of pointwise rerankers makes it difficult to calibrate relevance~\cite{xian2023learning}. Recent works~\cite{ma2023zero, sun2023chatgpt} adopted listwise reranking to generate the ordered list of candidates directly, without needing any intermediate relevance scores. Compared to pointwise or pairwise counterparts~\cite{qin2023large}, listwise reranking requires fewer runs as it takes multiple documents into account for a single window. When reranking multiple candidates making the prompt size more than max allowed input context length, listwise reranking adopts a sliding window strategy~\cite{sun2023chatgpt} with a fixed window and step size. However, due to the computationally demanding nature of LLMs, the improved results from listwise reranking come at the expense of increased latency.
Recent work has tackled the latency problem of listwise reranking through efficient processing of candidate passages.~\citet{meng2024rlt} introduced ranked list truncation, which optimizes the process by trimming reranking candidates, allowing for variable-length candidate lists that can be adapted per query. ~\citet{parry2024topdown} propose top-down partitioning, which introduces a parallelizable algorithm that effectively reduces redundancy in inference calls. Our method, \name{}, reduces the latency for each window in listwise reranking by lowering the number of output tokens required to be generated to one. \name{} complements existing strategies like ranked list truncation and top-down partitioning as each method targets a distinct yet complementary aspect of the listwise reranking workflow. We leave the empirical investigation of stacking these approaches together as an important direction for future work.




\section{Methodology}
\label{sec:methodology}
In this section, we first discuss the fundamentals of listwise LLM reranking (\S{\ref{sec:background}}). 
We then present \name{}, our own novel approach to the task (\S{\ref{sec:first_method}}).

\subsection{Listwise Reranking with LLMs}
\label{sec:background}
Given a list of retrieved passages $\mathcal{P} = \{p_1, p_2, ..., p_n\}$, the task of a reranker is to return $k$ passages that are the most relevant to a query $q$.
Due to input size limits, listwise reranking with LLMs often adopts a sliding window strategy with a window size of $m$ passages ($m < n$) and a step size $s$~\cite{sun2023chatgpt}.
For each window, passages are denoted by unique identifiers $t_i$; the LLM reranker generates as output a sequence of identifiers in decreasing order of their relevance (e.g., $t_1 > t_3 > t_2$).
The global process operates by first ranking the last $m$ documents and then iteratively sliding the processing window $s$ positions at a time until the beginning of the list is reached~\cite{sun2023chatgpt}.

Recent work~\cite{pradeep2023rankvicuna, pradeep2023rankzephyr} has drawn supervision for open-source listwise LLM rerankers~\cite{tunstall2023zephyr} from larger proprietary models, such as GPT$_{3.5}$ and GPT$_4$.
The relevance supervision in such cases comes in the form of a generated sequence $y = [y_1]>[y_2]...>[y_m]$, where $y_i$ is the identifier of a document that has been judged more relevant to the query $q$ than $y_j$, for every $m \geq j > i$.
The reranker is then trained with a language modeling objective, minimizing the error in predicting the true next token in the generation sequence:
\begin{equation}
    \mathcal{L}_{LM} = -\sum_{i=1}^{|y|}\text{log}(P_{\theta}(y_i|x,y_{<i}))
\end{equation}
$P_{\theta}(y_i|x,y_{<i})$ here is the conditional probability of predicting the target $y_i$ given the instruction prompt $x$ and the preceding tokens $y_{<i}$.

\begin{table*}[t]
    \centering
    \setlength{\tabcolsep}{0.3em}
    \def\arraystretch{1.7}
    \small
    \begin{tabular}{cccccccccccccc}
    \hline
    \multirow{2}{*}{\parbox{4em}{\centering\textbf{Reranker}}} & \multirow{2}{*}{\parbox{3.5em}{\centering\textbf{Training Data}}} & \multirow{2}{*}{\parbox{2em}{\centering\textbf{Avg.}}} & \multirow{2}{*}{\parbox{3em}{\centering Climate FEVER}} & \multirow{2}{*}{\parbox{2.75em}{\centering DBP- edia}}& \multirow{2}{*}{\parbox{3.25em}{\centering FEVER}} & \multirow{2}{*}{\parbox{2.25em}{\centering FiQA}} & \multirow{2}{*}{\parbox{2.75em}{\centering Hotpot QA}}& \multirow{2}{*}{\parbox{2.25em}{\centering MS Marco}} & \multirow{2}{*}{\parbox{2.5em}{\centering NFC- orpus}} & \multirow{2}{*}{\parbox{1.5em}{\centering NQ}} & \multirow{2}{*}{\parbox{2.25em}{\centering Sci- docs}} & \multirow{2}{*}{\parbox{2.25em}{\centering Sci- fact}} & \multirow{2}{*}{\parbox{3em}{\centering Trec-COVID}} \\ \\
    \hline
    None & MS Marco & 45.9 & 23.7 & 41.3 & 75.8 & 32.9 & 63.8 & 40.7 & 32.8 & 49.8 & 16.5 & 67.7 & 59.6 \\
    Cross-Encoder & MS Marco & 50.7 & 25.5 & 47.0 & \textbf{81.9} & 35.6 & 71.8 & \textbf{47.0} & 34.5 & 57.6 & 17.0 & 69.1 & 71.0 \\
    Rank Vicuna & GPT 3.5 & 50.7 & \textbf{28.2} & 50.0 & 81.0 & 35.9 & 73.5 & 36.7 & 33.1 & 58.6 & 18.4 & 70.5 & 71.3\\
    Rank Zephyr & GPT 3.5+4 & 53.7 & 25.6 & 50.0 & 80.1 & \textbf{42.2} & 71.6 & 42.7 & \textbf{37.7} & 65.6 & \textbf{20.5} & \textbf{76.7} & 78.4\\
    \hline
    \name{} & GPT-4 & \textbf{54.3} & 26.7 & \textbf{50.9} & 81.7 & \textbf{42.2} & \textbf{74.2} & 44.4 & 37.4 & \textbf{66.4} & 20.4 & 74.6 & \textbf{78.8} \\
    \hline
    \end{tabular}
    \caption{
    Performances of different rerankers (nDCG@10 in \%) on BEIR~\cite{thakur2021beir}.
    Top-100 retrieval results from Contriever~\cite{gautier2022unsupervised} are passed as input. 
    Reranker: \textit{None} indicates the retriever.
    }
    \label{tab:ranking_performance}
\end{table*}

\subsection{\name{}: Ranking with a Single Token}
\label{sec:first_method}
The \name{} method operates under the hypothesis -- which we validated in \S{\ref{sec:introduction}} -- that LLMs can latently approximate the full ranked list during the generation of the first (top-ranked) passage identifier.
\name{} simply extracts the output logits of candidate identifier tokens while generating the first identifier $y_1$ and returns the passage ranking in the order of decreasing logit values.
Crucially, this process only involves computing the output logits of a single token during inference.

Since this ranking is based on output logits of individual tokens from the LLM's vocabulary, avoiding tokenizing passage identifiers into multiple tokens is key.
Using numeric identifiers would limit the number of candidates to $\leq 9$ as byte pair encoding ~\cite{sennrich2016bpe} tokenizes multiple-digit numbers into more than one token.
We, therefore, adopt alphabetic identifiers instead, ranging from A to Z, as LLM rerankers typically consider up to 20 candidate passages in a single window.

Using \name{} directly with current LLM rerankers~\cite{pradeep2023rankvicuna, pradeep2023rankzephyr}, while showing promise in the evaluation of Figure~\ref{fig:ranking_matching}, is still suboptimal, as these models are finetuned with a language modeling objective.
Hence, we propose to leverage a learning-to-rank objective to provide targeted supervision to \name{} rerankers that can better equip them to rank using the first token's output logits. 
Formally, given $m$ candidate passages $(p_1, p_2, ..., p_m)$, with $t_i$ the identifier token of $p_i$ and $s_i$ the output vocabulary logit of passage identifier $t_i$ during first token generation, let $r_i \in [1, 2, ..., m]$ be the true rank of $p_i$ within the $m$ candidates.
We consider as our training objective a weighted version of RankNet~\cite{burges2005ranknet} -- a pairwise loss which considers the correctness of relative passage orders to formulate the learning-to-rank objective -- as follows:
\begin{equation}
\label{eq:weighted_ranknet}
\begin{aligned}
    \mathcal{L}_{Rank} & = \sum_{i=1}^m\sum_{j=1}^m \frac{\mathds{1}_{r_i < r_j}}{i + j} \text{log}(1+ \text{exp}(s_i - s_j)) \\
    & = \sum_{r_i < r_j} \frac{1}{i + j} \text{log}(1+ \text{exp}(s_i - s_j))
\end{aligned}
\end{equation}
Here, the weight $1/(i+j)$ is the inverse mean rank of candidate pair $(i, j)$, which prioritizes getting the ranks of higher-ranked candidates right over those of lower-ranked ones.
Since the 
standard
language modeling objective has also 
been used successfully to train listwise 
rerankers,
we combine it with $\mathcal{L}_{Rank}$ to construct the following joint loss for our training:
\begin{equation}
\label{eq:joint_loss}
    \mathcal{L}_{Joint} = \mathcal{L}_{LM} + \lambda \mathcal{L}_{Rank}
\end{equation}
where $\lambda$ is a hyperparameter 
that controls the relative importance of the two losses.
Note that while $\mathcal{L}_{Rank}$ is applied only to the output logits of the first generated token, $\mathcal{L}_{LM}$ is 
an aggregate over all tokens in the target ranking sequence.
At inference, \name{} uses only the output vocabulary logits of the first generation token to obtain the ranked candidate identifier order.



\section{Experiments}
\label{sec:experiments}

\begin{table*}[t]
    \centering
    \setlength{\tabcolsep}{0.3em}
    \def\arraystretch{1.7}
    \small
    \begin{tabular}{cccccccccccccc}
    \hline
    \multirow{2}{*}{\parbox{4em}{\centering\textbf{Training Strategy}}} & \multirow{2}{*}{\parbox{3.5em}{\centering\textbf{Inference}}} & \multirow{2}{*}{\parbox{2.25em}{\centering\textbf{Avg.}}} & \multirow{2}{*}{\parbox{3em}{\centering Climate FEVER}} & \multirow{2}{*}{\parbox{2.75em}{\centering DBP- edia}}& \multirow{2}{*}{\parbox{3.25em}{\centering FEVER}} & \multirow{2}{*}{\parbox{2.25em}{\centering FiQA}} & \multirow{2}{*}{\parbox{2.75em}{\centering Hotpot QA}}& \multirow{2}{*}{\parbox{2.25em}{\centering MS Marco}} & \multirow{2}{*}{\parbox{2.5em}{\centering NFC- orpus}} & \multirow{2}{*}{\parbox{1.5em}{\centering NQ}} & \multirow{2}{*}{\parbox{2.25em}{\centering Sci- docs}} & \multirow{2}{*}{\parbox{2.25em}{\centering Sci- fact}} & \multirow{2}{*}{\parbox{3em}{\centering Trec-COVID}} \\ \\
    \hline
    LM & Generation & 52.3 & 20.8 & 48.6 & 79.1 & 40.6 & 71.3 & 43.5 & 35.4 & 65.6 & 19.7 & 72.3 & 77.6 \\
    \hline
    LM+RankNet & \multirow{3}{*}{\parbox{3em}{\centering \name{}}} & \textbf{54.3} & \textbf{26.7} & \textbf{50.9} & \textbf{81.7} & 42.2 & 74.2 & 44.4 & \textbf{37.4} & 66.4 & \textbf{20.4} & 74.6 & \textbf{78.8} \\
    - Weighting &  & 53.8 & 23.7 & 50.1 & 79.0 & \textbf{43.2} & \textbf{74.9} & \textbf{44.6} & 36.8 & \textbf{66.9} & 19.7 & \textbf{75.3} & 77.5 \\    
    - LM & & 51.7 & 20.3 & 48.8 & 74.8 & 40.6 & 72.5 & 43.2 & 35.9 & 63.5 & 19.3 & 73.5 & 76.2 \\
    \hline
    \end{tabular}
    \caption{Table showing the nDCG@10 (in \%) on BEIR~\cite{thakur2021beir} for LLM listwise reranking when training with different strategies. \textit{LM} corresponds to the traditional language modeling objective for training.}
    \label{tab:ranking_ablation}
\end{table*}

We first demonstrate in \S{\ref{sec:ranking}} that the proposed ranking loss improves the accuracy of listwise LLM reranking.
Next, in \S{\ref{sec:latency}}, we measure the improvement in latency of inference from using \name{}. 
Finally, we show in \S{\ref{sec:relevance_feedback}} that leveraging listwise LLM rerankers for relevance feedback improves the recall of retrievers.

\subsection{Setup}

\noindent \paragraph{Model:} 
We follow~\citet{pradeep2023rankzephyr} to use Zephyr$_{\beta}$~\cite{tunstall2023zephyr} as our instruction-following LLM for listwise reranking.
Zephyr$_{\beta}$ is a 7B LLM based on Mistral~\cite{jiang2023mistral} and instruction-tuned on chat datasets~\cite{ding2023enhancing, cui2023ultrafeedback}.
We finetune Zephyr$_{\beta}$ for listwise reranking for three epochs with an effective batch size of 32, a learning rate of 5e-6 using bfloat16 precision, and leverage noisy embeddings~\cite{jain2023neftune}.
Training takes approximately 7 hours on four 40GB Nvidia A100 GPUs when used with DeepSpeed~\cite{rasley2020deepspeed}. We randomly sample 300 queries from MS Marco as our development set, and use $\lambda=10$ for scaling the weighted RankNet loss

\noindent \paragraph{Datasets:} 
We use 40k GPT-4 labeled instances from~\citet{pradeep2023rankzephyr} for fine-tuning LLM rerankers, which were created using 5k queries from MS MARCO~\cite{nguyen2016ms}.
Examples contain a variable number ($\le20$) of candidate passages that need to be reranked. 
For evaluation, we use the BEIR benchmark~\cite{thakur2021beir}, which comprises test instances from MS MARCO and out-of-domain evaluation data from several scientific, biomedical, financial, and Wikipedia-based retrieval datasets\footnote{We use the same BEIR subset as in~\citet{reddy2023inference}.}.

\noindent \paragraph{Reranking Setup:} 
We use Contriever~\cite{gautier2022unsupervised} for retrieving an initial set of candidates.
The top 100 retrieved passages are then passed as input to the reranker. The listwise reranking process uses a sliding window strategy as in ~\citet{sun2023chatgpt, pradeep2023rankzephyr}, with window size $m=20$ and step size $s=10$.

\noindent \paragraph{Baselines:}
We compare performance with a pointwise cross-encoder reranker from ~\citet{thakur-2020-AugSBERT}, as well as RankVicuna~\cite{pradeep2023rankvicuna} and RankZephyr~\cite{pradeep2023rankzephyr}, which are LLM-based listwise rerankers.
The cross-encoder was trained using 500k pairwise human-annotated instances from MS 
MARCO~\cite{nguyen2016msmarco}. RankVicuna was finetuned using the RankGPT data~\cite{sun2023chatgpt}, which contains GPT-3.5 
labeled
listwise reranking examples created from 100k MS 
MARCO
queries. 
RankZephyr employs a two-stage training process that first finetunes with the RankGPT data and then with GPT-4 labeled listwise reranking examples created from 5k MS MARCO queries.
We only use the smaller GPT-4 labeled instances due to compute constraints.

\subsection{Ranking Performance}
\label{sec:ranking} 
Table \ref{tab:ranking_performance} shows nDCG@10 scores of different rerankers on BEIR~\cite{thakur2021beir}, where each reranker was used to rerank the top-100 retrievals of Contriever.
We first observe that \name{} outperforms RankZephyr despite being fine-tuned on considerably less data.
Note that the cross-encoder achieves a very high score on MS MARCO as it was trained with in-domain human-annotated data, unlike the LLM rerankers.

Next, we report results from ablation studies involving different finetuning strategies in Table \ref{tab:ranking_ablation}.
The proposed joint loss significantly improves performance over finetuning with just the language modeling objective.
The benefit of adding the proposed inverse mean rank weighting to the existing RankNet loss is also evident.
Interestingly, we observe that finetuning using only the weighted RankNet loss performs worse than using only the LM objective, which is perhaps unsurprising given the alignment of the latter with LLM pretraining.

Further, in addition to the weighted RankNet loss (eq. \ref{eq:weighted_ranknet}), we experimented with incorporating different ranking losses while finetuning the listwise reranker. Specifically, we considered the LambdaRank and ListNet losses. LambdaRank~\cite{burges2006lambdarank} is a pair-wise ranking loss that is similar to RankNet, but uses a weight proportional to the change in the target ranking metric (e.g. NDCG) that would result from swapping the positions of items in the pair. ListNet~\cite{cao2007listnet} is a listwise loss based on the cross entropy between two parameterized probability distributions of permutations. Table \ref{tab:ranking_loss_comparison} shows the results on a subset of BEIR. We see that our weighted RankNet loss gives a better performance compared to using the LambdaRank and ListNet losses. 

\begin{figure*}[t]
    \centering
    \begin{subfigure}{0.33\linewidth}       
     \includegraphics[scale=0.52]{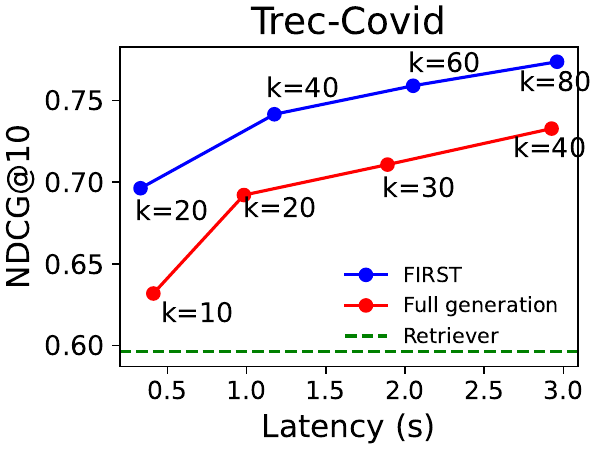}
     \vspace{-0.5em}
     \end{subfigure}%
     \hfill
    \begin{subfigure}{0.33\linewidth}       
     \includegraphics[scale=0.52]{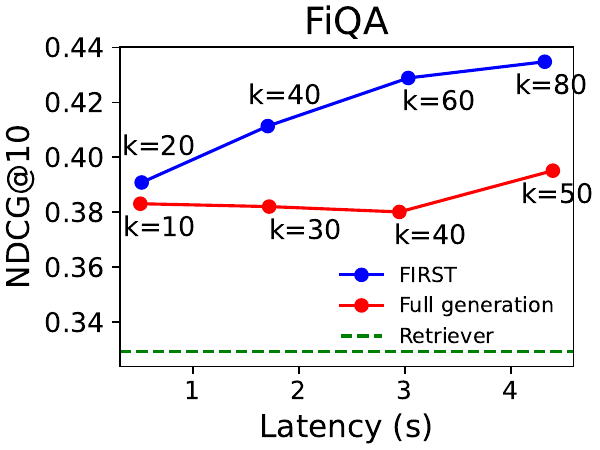}
     \vspace{-0.5em}
     \end{subfigure}%
     \hfill
     \begin{subfigure}{0.33\linewidth}       
     \includegraphics[scale=0.52]{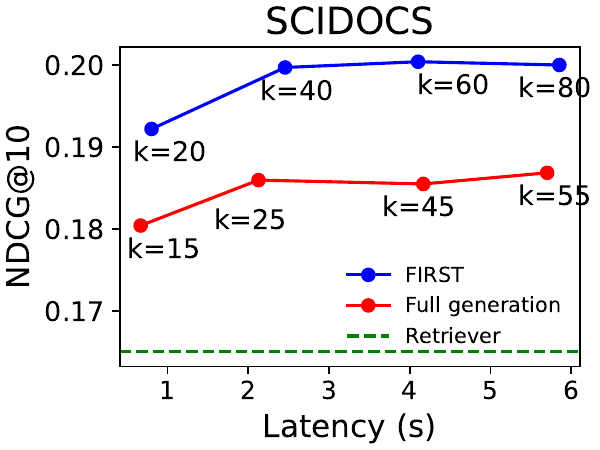}
     \vspace{-0.5em}
     \end{subfigure}%
     \hfill
     \begin{subfigure}{0.33\linewidth}       
     \includegraphics[scale=0.52]{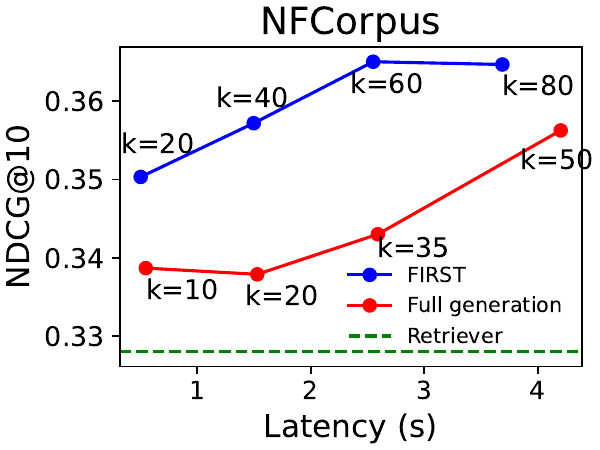}
     \end{subfigure}%
     \hfill
    \begin{subfigure}{0.33\linewidth}       
     \includegraphics[scale=0.52]{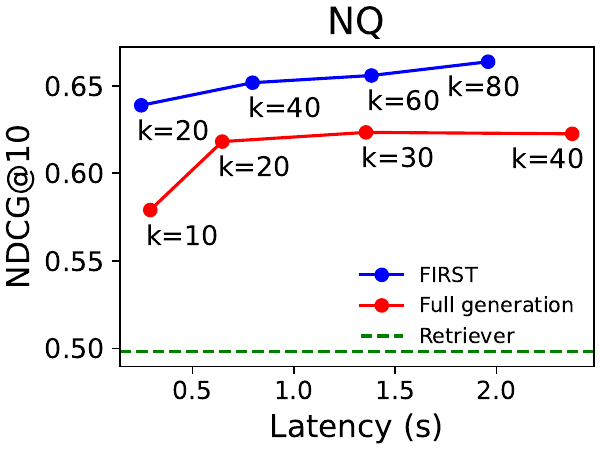}
     \end{subfigure}%
     \hfill
     \begin{subfigure}{0.33\linewidth}       
     \includegraphics[scale=0.52]{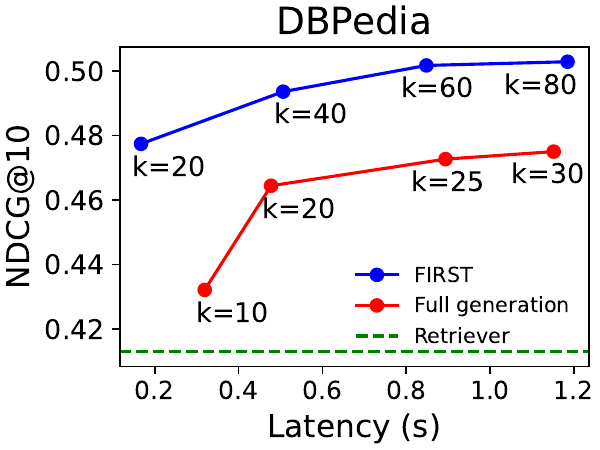}
     \end{subfigure}%
    \caption{
    Ranking accuracy (nDCG@10) against the reranker's per query latency in seconds.
    $k$ refers to the number of passages reranked for the corresponding latency.
    \name{} considerably outperforms sequence generation when constrained to a latency budget, as it is able to rerank significantly more candidates.}
    \label{fig:performance_vs_latency}
\end{figure*}

\begin{table}[t]
    \centering
    \setlength{\tabcolsep}{0.5em}
    \def\arraystretch{1.5}
    \small
    \begin{tabular}{c|c|c|c}
    \textbf{Dataset} & \textbf{RankNet} & \textbf{LambdaRank} & \textbf{ListNet} \\
    \hline
    DBPedia & \textbf{50.9} & 47.3 & 49.1\\
    FiQA & 42.2 & 43.2 & \textbf{43.7}\\
    NFCorpus & \textbf{37.4} & 35.6 & 36.8 \\
    Scifact & 74.6 & \textbf{76.1} & 74.4 \\
    Trec-COVID & \textbf{78.8} & 75.0 & 75.5\\
    \hline
    Average & \textbf{56.7} & 55.4 & 55.9 \\
    \hline
    \end{tabular}
    \caption{Table showing the nDCG@10 (in \%) on a subset of BEIR from incorporating different ranking losses when finetuning the listwise LLM reranker.}
    \label{tab:ranking_loss_comparison}
\end{table}

\begin{figure}[t]
    \centering
    \includegraphics[scale=0.7]{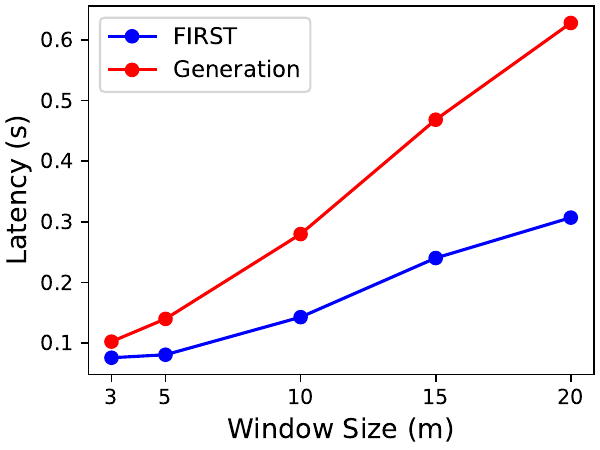}
    \caption{Plot comparing the single window inference latency for \name{} vs. generating the ranked sequence, for different numbers of candidate passages $m$.}
    \label{fig:onepass_latency}
    \vspace{-1em}
\end{figure}

\subsection{Comparing Latencies}
\label{sec:latency}

One of the key stated advantages of \name{} is single-token decoding, which can be expected to improve latency considerably.
To demonstrate this empirically, we compare the latencies of inference with \name{} and sequence generation\footnote{For a fair comparison, we omitted the generation time of the identifier indicators (`[' and `]') for sequence generation.}.
Latency is measured on a 40GB Nvidia A100 GPU and averaged over 200 sampled queries.

We first compare the overall time taken for ranking candidate passages in a single window.
Figure \ref{fig:onepass_latency} plots the latency 
of
\name{} and sequence generation against the window size $m$. 
While overall inference time increases for both approaches with more candidate passages in the window, the latency gap between the two grows as $m$ increases. 
This is understandable, as the output length increases for sequence generation with the number of candidate passage identifiers, but not for \name{}.


In Figure \ref{fig:performance_vs_latency}, we further evaluate the reranking accuracy of the two approaches under specific latency requirements.
We fix the number of the candidates $k = (20, 40, 60, 80)$ for \name{} and retrieve the corresponding number of candidates with sequence generation under identical latency requirements.
Figure \ref{fig:performance_vs_latency} shows the plots for six different datasets from BEIR, where we observe \name{} to consistently outperform sequence generation while maintaining the same per-query reranking latency.
Clearly, \name{} can rerank more candidates $k$ in the same amount of time, which leads to the observed performance gains.

\subsection{Relevance Feedback with LLM Rerankers}
\label{sec:relevance_feedback}

\begin{table*}[t]
    \centering
    \setlength{\tabcolsep}{0.3em}
    \def\arraystretch{1.7}
    \small
    \begin{tabular}{p{8.5em}|cccccccccccc}
    \hline
    \multirow{2}{*}{\parbox{8.5em}{\centering\textbf{Relevance Feedback}}} & \multirow{2}{*}{\parbox{4em}{\centering\textbf{Average R@100}}} & \multirow{2}{*}{\parbox{3em}{\centering Climate FEVER}} & \multirow{2}{*}{\parbox{2.75em}{\centering DBP- edia}}& \multirow{2}{*}{\parbox{3.25em}{\centering FEVER}} & \multirow{2}{*}{\parbox{2.25em}{\centering FiQA}} & \multirow{2}{*}{\parbox{2.75em}{\centering Hotpot QA}}& \multirow{2}{*}{\parbox{2.25em}{\centering MS Marco}} & \multirow{2}{*}{\parbox{2.5em}{\centering NFC- orpus}} & \multirow{2}{*}{\parbox{1.5em}{\centering NQ}} & \multirow{2}{*}{\parbox{2.25em}{\centering Sci- docs}} & \multirow{2}{*}{\parbox{2.25em}{\centering Sci- fact}} & \multirow{2}{*}{\parbox{3em}{\centering Trec-COVID}} \\ \\
    \hline
    \centering None & 66.8 & 57.4 & 54.1 & 94.9 & 65.6 & 77.7 & 89.1 & 30.0 & 92.5 & 37.8 & 94.7 & 40.7 \\
    \centering CE (KL Div.) & 69.0 & \textbf{59.5} & 57.3 & \textbf{95.5} & 65.6 & 80.4 & \textbf{90.5} & 31.9 & 94.2 & 40.1 & 95.2 & 51.5 \\
    \centering LLM (RankNet) & 71.2 & 58.8 & 58.4 & 95.2 & \textbf{72.7} & 79.8 & 89.3 & 34.5 & 95.6 & 43.1 & \textbf{96.1} & 59.4 \\
    \hline
    \centering CE + LLM & \textbf{72.0}  & 59.4 & \textbf{59.8} & \textbf{95.5} & 71.8 & \textbf{81.2} & 89.7 & \textbf{35.9} & \textbf{96.1} & \textbf{44.1} & 95.9 & \textbf{62.2}\\
    \hline
    \end{tabular}
    \caption{Table showing recall@100 (in \%) on BEIR~\cite{thakur2021beir} using the updated query vector for second-stage retrieval after relevance feedback. Results for \textit{None} correspond to the first-stage retrieval using Contriever. Relevance feedback from cross-encoder (CE) uses the KL divergence loss as in~\citet{reddy2023inference}, while that from listwise LLM reranker uses the weighted RankNet loss (Eq. \ref{eq:weighted_ranknet}) for optimizing the query vector.}
    \label{tab:relevance_feedback}
\end{table*}

Here, we demonstrate that the better ranking performance from LLM-based rerankers, when compared to cross-encoders, is advantageous for downstream applications. Specifically, we consider the task of providing relevance feedback~\cite{rocchio1971relevance} for improving the retrieval recall.
Relevance feedback using rerankers at inference involves optimizing the retriever's query representation at test-time using the reranker's output for the retrieval results.  ~\citet{reddy2023inference, sung2023optimizing} update the query representation from dense retrievers, like Contriever~\cite{gautier2022unsupervised}, by gradient descent based on KL divergence loss between the query vector and cross-encoder reranker scoring distributions over the retrieved passages. Since rerankers are typically more performant than retrievers, the updated query representation, when used for second-stage retrieval, can improve recall upon the previously retrieved results. We refer the reader to~\citet{reddy2023inference} for more details.

While cross-encoder rerankers provide floating-point scores that can be used as distillation supervision, listwise rerankers output an ordered sequence of the candidates. Hence, the typically used KL divergence loss cannot be applied for relevance feedback in this setting.  In this regard, we investigate how listwise rerankers can be leveraged for relevance feedback, and whether they can provide bigger improvements for second-stage retrieval recall compared to cross-encoders. We experiment with using the weighted RankNet loss (in eq. \ref{eq:weighted_ranknet}) to use the ranked ordering from listwise rerankers as distillation supervision for relevance feedback. 

For our experiments, we follow the same setup as~\citet{reddy2023inference} with Contriever for initial retrieval and evaluation on BEIR~\cite{thakur2021beir}. Distillation using the cross-encoder with KL divergence loss has a learning rate of $0.005$ and $100$ gradient updates, while that using the LLM reranker with the weighted RankNet loss has a learning rate of $0.001$ and $20$ gradient updates. Table \ref{tab:relevance_feedback} shows recall@100 numbers from second-stage retrieval after different relevance feedback strategies. We observe that relevance feedback from the LLM reranker significantly improves recall compared to the cross-encoder reranker. We attribute this to the superior ranking performance of LLM rerankers (as seen in Table \ref{tab:ranking_performance}), thereby providing higher quality relevance feedback. Moreover, we see that using the LLM reranker feedback in addition to that from the cross-encoder (CE+LLM) leads to further gains. This improvement could be explained as the diversity of feedback signals from the two rerankers, i.e. floating-point scores for cross-encoder vs ranking sequence for listwise reranker, providing a more comprehensive distillation supervision and demonstrating the huge potential of listwise rerankers for relevance feedback.


\section{Conclusion}

In this work, we introduce \name{}, a novel strategy for listwise LLM reranking. \name{} leverages the output logits of the first generated identifier to obtain a ranking for the candidates, as opposed to the typical approach of generating the entire ranked ordering sequence of candidate passage identifiers. We demonstrated that our single-token decoding approach reranks a considerably larger number of candidates compared to inference with ordered sequence generation in the same time, leading to larger gains when reranking under a latency constraint. \name{} also demonstrates ranking performance benefits from incorporating a learning-to-rank loss during training, allowing for prioritizing more important ranks. By addressing both the training and inference inefficiencies of existing LLM listwise reranking approaches, \name{} represents a significant step forward in the development of advanced re-ranking techniques using LLMs.

\section*{Limitations}

While \name{} benefits from leveraging GPT-4 labeled data for training, we have not experimented with using human-annotated pairwise examples in supervised datasets such as MS Marco to further improve performance. Moreover, our experiments here are on English data on account of the underlying LLM being predominantly monolingual. An interesting extension would be to finetune a multilingual LLM for listwise reranking to demonstrate the benefit of our approach in other languages. Further, we use alphabets as passage identifiers since the window size for listwise reranking is typically $\le$20. However, we expect finetuning using other vocabulary tokens as identifiers should enable leveraging a larger set of candidate identifiers in case the window size needs to be further increased. 

\section*{Acknowledgements}
We acknowledge Ron Arel, Rishub Tamirisa and Andy Zhou from Lapis Labs for helping with access to NCSA compute. We would also like to thank members of the BlenderNLP group for valuable comments and feedback. We are grateful to Ronak Pradeep for releasing the training data and code for RankZephyr. This research is based on work supported by U.S. DARPA KAIROS Program No. FA8750-19-2-1004, and the Molecule Maker Lab Institute: an AI research institute program supported by NSF under award No. 2019897 and No. 2034562. The views and conclusions contained herein are those of the authors and should not be interpreted as necessarily representing the official policies, either expressed or implied, of DARPA, or the U.S. Government. The U.S. Government is authorized to reproduce and distribute reprints for governmental purposes notwithstanding any copyright annotation therein.



\bibliography{custom}

\end{document}